\newcommand {\be}{\begin{equation}}
\newcommand {\ee}{\end{equation}}

\newcommand {\bea}{\begin{eqnarray}}
\newcommand {\eea}{\end{eqnarray}}

\documentstyle[aps,prb]{revtex}
\draft

\begin{document}

\title{
Dynamics of the spin-half Heisenberg chain at intermediate temperatures}
\author{O. A. Starykh,$^{1,}$\cite{byline1,byline2} A. W. Sandvik$^{2,}$ 
\cite{byline3} and R. R. P. Singh$^{1}$}
\address{$^{1}$Department of Physics, University of California, Davis, 
California 95616\\
$^{2}$National High Magnetic Field Laboratory, Florida State University, 
1800 East Paul Dirac Drive, Tallahasse, FL 32306}
\date{\today}
\maketitle{}

\begin{abstract}
Combining high-temperature expansions with the recursion method and quantum 
Monte Carlo simulations with the maximum entropy method, we study the 
dynamics of the spin-$1/2$ Heisenberg chain at temperatures above and 
below the coupling $J$. By comparing the two sets of calculations, their 
relative strengths are assessed. At high temperatures, we find that there is 
a low-frequency peak in the momentum integrated dynamic structure factor,
due to diffusive long-wavelength modes. This peak is rapidly suppressed as the 
temperature is lowered below $J$. Calculation of the complete dynamic 
structure factor $S(k,\omega)$ shows how the spectral features associated
with the two-spinon continuum develop at low temperatures. We extract the 
nuclear spin-lattice relaxation rate $1/T_1$ from the $\omega \to 0$ limit, 
and compare with recent experimental results for Sr$_2$CuO$_3$ and
CuGeO$_{\rm 3}$. We also discuss the scaling behavior of the dynamic 
susceptibility, and of the static structure factor $S(k)$ and the static 
susceptibility $\chi (k)$. We confirm the asymptotic low-temperature forms 
$S(\pi)\sim [\ln{(T)}]^{3/2}$  and $\chi(\pi)\sim T^{-1}[\ln{(T)}]^{1/2}$, 
expected from previous theoretical studies.
\end{abstract}

\section{Introduction}
Quantum antiferromagnets represent an important class of systems in both 
theoretical and experimental condensed matter physics. In recent years,
greatly improved precision of neutron scattering and NMR experiments have 
made possible very detailed comparisons with theoretical predictions. A number
of new materials have been synthesized which appear to be near-perfect 
realizations of the simple spin-half Heisenberg model in various geometries. 
For example, Sr$_{\rm 2}$CuO$_{\rm 3}$,\cite{Srchain,Takigawa}
SrCu$_{\rm 2}$O$_{\rm 3}$,\cite{Srladder} and 
Sr$_{\rm 2}$CuO$_{\rm 2}$Cl$_{\rm 2}$,\cite{Srlayer} 
comprise structural copper-oxygen units with magnetic properties well 
described by the Heisenberg model on a single 1D chain, two coupled chains, 
and a 2D square lattice, respectively. Sr$_{\rm 2}$CuO$_{\rm 3}$ is interesting
because it appears to be the most perfect 1D spin-half Heisenberg system found
so far, with an exchange $J \approx 2000$K and a 3D ordering temperature 
$T_{\rm N} \approx 5K$. Detailed experimental studies of this system, 
\cite{Srchain,Takigawa} as well as other quasi-1D materials such as 
KCuF$_{\rm 3}$ \cite{Tennant} and CuGeO$_{\rm 3}$,\cite{CuGeO} have pointed 
to the need for more accurate theoretical studies of the spin dynamics of 
the $S=1/2$ Heisenberg chain. Although this model, defined in standard 
notation by the Hamiltonian
\be
H=J\sum_i {\bf S}_i \cdot {\bf S}_{i+1},\quad (J > 0),
\label{hamilton}
\ee
is perhaps the most studied of the basic interacting quantum many-body 
models, its finite-temperature dynamic properties are not fully understood.
The low-temperature ($T \ll J$) behavior is controlled by 
the $T=0$ quantum critical point (line of critical points to be 
exact). The powerful machinery of bosonization and conformal field theory 
enables one to make a number of experimentally verifiable predictions in 
this regime.\cite{Schulz,Sachdev} The high temperature regime ($T \gg J$) 
has been studied numerically by short-time or frequency moment expansions 
combined with the recursion method.\cite{Muller} The regime of 
intermediate temperatures $T\approx J$ is the most difficult to study 
theoretically, but is clearly of much experimental and theoretical 
significance, containing the crossover from the diffusive high-temperature 
behavior to the low-temperature regime dominated by elementary-excitations. 
Here we study the dynamics at intermediate temperatures using the 
high-temperature expansion (HTE) technique and a recently developed 
``stochastic series expansion'' quantum Monte Carlo (QMC) technique
\cite{Sandvik1,Elescorial} (an improved variant of the so called Handscomb's 
method \cite{Handscomb}). We have also numerically diagonalized the 
Hamiltonian for a chain with $16$ spins, which although not large enough
to give reliable results in general, provides for a good test of the other
methods in certain regimes. 

The HTE method has been extensively used to study static properties of 
spin-models.\cite{HTErefs} Here we combine it with the continued fraction 
(or recursion) method \cite{Muller} to calculate dynamic properties at finite 
temperatures. The QMC method used here has also previously been applied to 
both statics and dynamics of Heisenberg models in several different 
geometries. \cite{SSEwork,Sandvik2} Accurate results for imaginary-time 
dependent correlation functions can be obtained down to fairly low 
temperatures. The maximum-entropy (Max-Ent) method \cite{maxent1,maxent2} 
is used for analytic continuation to real frequencies. This approach has 
previously been applied to the spin dynamics of the 1D Heisenberg model by 
Deisz, Jarrell and Cox,\cite{Deisz1,Deisz2} who focused mainly on the 
low-temperature dynamic structure factor and the differences between 
half-integer and integer spin. They also discussed at length the accuracy 
of the Max-Ent method. Here we find that for static properties, results 
obtained using HTE and QMC agree almost perfectly for $T/J \agt 1/8$, below 
which the HTE method becomes unreliable. For dynamic properties, the HTE 
method performs well for $T/J \agt 0.5$, and in this regime the results 
agree well with those of QMC and the Max-Ent method. This gives us confidence
that the QMC and Max-Ent techniques are reliable at lower temperatures  
as well.

Our main results are the following: At high temperatures, we find that the 
dynamic structure factor $S(k,\omega)$ grows sharply (perhaps diverges) as
$k \to 0$ and $\omega \to 0$. This diffusive behavior leads to a low-frequency
peak in the momentum average
\be
S_A(\omega)=\int {dk \over 2\pi} |A(k)|^2S(k,\omega),
\label{sa}
\ee
if the form factor $A(k=0)$ is non-zero. In an NMR experiment, depending on 
the $A(k)$ corresponding to a given material and nucleus under study, this 
can have large effects on the spin-lattice relaxation rate, which is given 
by \cite{Moriya}
\be
{1 \over T_1} = 2S_A(\omega \to 0).
\label{t1def}
\ee
QMC+Max-Ent results for $1/T_1$ were previously reported in
Ref.~\onlinecite{Sandvik2}. Here we provide results of higher accuracy, 
obtained by calculating the full momentum dependence of $S(k,\omega \to 0)$. 

As the temperature is lowered below $T \approx J/2$, the diffusive peak 
rapidly diminishes in magnitude, and the low-frequency spectral weight shifts 
to $k=\pi$, as expected. The QMC results for $S(k,\omega)$ clearly show the 
emergence of spectral features that can be associated with the well known 
\cite{Muller2,Schulz,Sachdev} $T=0$ two-spinon continuum. 

Comparisons of the momentum- and frequency-dependent numerical data with 
scaling theories at low temperatures have been presented elsewhere.\cite{SSS} 
Here we briefly discuss how the scaling in $q/T$ is violated due to 
logarithmic corrections. We provide numerical results for the temperature
dependence of the staggered structure factor $S(\pi)$ and the static staggered
susceptibility $\chi(\pi)$. At low temperatures the former behaves as 
$[\ln{(1/T)}]^{3/2}$, while the latter behaves as $T^{-1}[\ln{(1/T)}]^{1/2}$, 
both expected from theoretical results.

In Sec.~II we discuss the dynamic structure factor and the computational 
methods used in this study. The results are presented in Sec.~III, and
in Sec.~IV we discuss and summarize our main conclusions.

\section{Basic Definitions and Numerical Techniques}

We begin by reviewing some basic definitions of the static and dynamic 
correlation functions we wish to calculate. 
Both the neutron scattering intensity 
and the NMR spin-lattice relaxation rate measure the dynamics of the 
electronic spin system through coupling via the operator $S^+_k$. Hence, 
the relevant dynamic correlation function is $\langle S^-_{-k}(t)S^+_k(0) 
\rangle$. In a spin-rotationally invariant system, which is considered here, 
this can be evaluated with respect to any quantization axis, and in numerical 
calculations it is most practical to choose the component diagonal in 
the representation chosen. Hence, we study the time-dependent spin-spin
correlation function
\begin{equation}
S_r (t)= \langle S^z_r (t) S^z_0 (0) \rangle,
\label{tcorr}
\end{equation}
were $S^z_r (t)$ denotes the $z$-component of a spin-$1/2$ operator at site 
$r$ at time $t$, and brackets denote thermodynamic averaging at temperature
$T/J=\beta^{-1}$. We consider only the case of zero average magnetization; 
$\langle S^z_r \rangle=0$. The dynamic structure factor $S(k,\omega)$ is 
the space-time Fourier transform of Eq.~(\ref{tcorr}):
\begin{equation}
S(k,\omega) = \sum\limits_{r} \int\limits_{-\infty}^\infty dt
{\rm e}^{-i(\omega t - kr)} 
\langle S^z_{r} (t) S^z_{0} (0) \rangle.
\label{sqwdef}
\end{equation}
Apart from kinematic factors, the neutron scattering cross section is 
directly proportional to $S(k,\omega)$. 

NMR (and related techniques such as NQR) can provide accurate results
for the low-frequency dynamics, through the spin-lattice relaxation rate,
given by Eqs.~(\ref{sa}) and (\ref{t1def}). The hyperfine form-factor
$A(k)$ can be obtained from the Knight shift, and also from impurity 
effects.\cite{Takigawa2,Eggert} Here we will restrict our attention to the 
important case where the nucleus under consideration resides on the sites 
of the electronic spins, and assume that the real-space hyperfine coupling 
$\bar A(r)$ has a on-site (direct) term $\bar A(0)$ and a nearest neighbor 
(transferred) term $\bar A(1)$, giving 
$A(k)=\bar A(0)+2\bar A(1)\cos{(k)}$. The spin-lattice relaxation rate is 
then given by 
\begin{equation}
\frac{1}{T_1}= 2\bar A^2(0)S_R(\omega = \omega_N),
\label{t1}
\end{equation}
where, $R=\bar A(1)/\bar A(0)$, $\omega_N$ is the resonance frequency, and we 
define
\be
S_R(\omega)= (1 + 2R^2)S_0(\omega) + 4R S_1(\omega) + 2R^2S_2(\omega),
\label{t11}
\end{equation}
where $S_r(\omega)$ is the real space dynamic spin correlation function
at distance $r$, i.e., the time Fourier transform of Eq.~(\ref{tcorr}).

The static structure factor $S(k)$ is the Fourier transform of the
equal-time correlation function, and the static susceptibility $\chi (k)$
is given by the Kubo integral
\be
\chi (k) = \sum\limits_{r} {\rm e}^{ikr} \int\limits_0^\beta d\tau
\langle S^z_{r} (\tau) S^z_{0} (0) \rangle,
\label{Kubo}
\ee
where $S^z_{r} (\tau) = {\rm e}^{\tau \hat H} S^z_{r} {\rm e}^{-\tau \hat H}$.
$S(k)$ and $\chi (k)$ can be related to the dynamic structure factor
through the sum rules \cite{Hohenberg}
\begin{mathletters}
\begin{eqnarray}
S(k)    & = & {1\over \pi} \int\limits_{0}^\infty d\omega 
(1+ {\rm e}^{-\beta\omega}) S(k,\omega), \\
\chi(k) & = & {2\over \pi} \int\limits_{0}^\infty d\omega \omega^{-1}
(1 - {\rm e}^{-\beta\omega}) S(k,\omega) .
\end{eqnarray}
\label{sumrules}
\end{mathletters}

Below we briefly describe the  numerical techniques we use to calculate 
the dynamic structure factor of the 1D Heisenberg model.

\subsection{High Temperature Expansion and the Recursion Method}

The correlation function (\ref{tcorr}) has a short-time expansion 
\begin{equation}
S_r(t)=\sum_{n=0}^{\infty} M_n \frac{(-i t)^n}{n!},
\label{shorttime}
\end{equation}
where the coefficients $M_n$ are defined as frequency moments,
\begin{equation}
M_n=\int_{-\infty}^{\infty} \frac{d\omega}{2\pi} \omega^n S_r(\omega),
\label{mn}
\end{equation}
of the time Fourier transform
\begin{equation}
S_r(\omega)=\int_{-\infty}^{\infty} dt {\rm e}^{-i\omega t} S_r(t).
\end{equation}
An important related function is the spectral density 
$\Phi(\omega)=(1 + e^{-\beta \omega})S(\omega)$, which is a real and
even function of frequency and, consequently, its inverse Fourier transform
$C_0(t)$ (often called the fluctuation function) has an expansion in even
powers of time only:
\begin{equation}
C_0(t)=\sum_{k=0}^{\infty} (-1)^k M_{2k} ~\frac{(-it)^{2k}}{(2k)!} .
\label{c0t}
\end{equation}
A short-time expansion is of little help if one is interested in the 
asymptotic long-time behavior of $C_0(t)$, unless some kind of analytic 
ansatz (most often a gaussian one) is made. To this end, let us define the 
relaxation function \cite{Muller}
\begin{equation}
c_0(z)=\int_0^{\infty} dt e^{-zt} C_0(t)=\sum_{k=0}^{\infty} M_{2k}
z^{-(2k + 1)}.
\label{c0z}
\end{equation}
From Eq.~(\ref{c0t}), one then has 
\begin{equation}
c_0(z)=\int_{-\infty}^{\infty}
\frac{d\omega}{2\pi} \frac{\Phi(\omega)}{z + i\omega},
\end{equation}
and $\Phi(\omega^{\prime})=2~\lim_{\epsilon \rightarrow +0}
Re [c_0(\epsilon - i\omega^{\prime})]$. Thus, upon analytic continuation
the relaxation function gives the spectral density. A useful property 
of the relaxation function is that it has a continued fraction 
representation: \cite{Muller}
\begin{equation}
c_0(z)=\frac{\Delta_0}{z + \frac{\Delta_1}{z + \frac{\Delta_2}{z + ...}}}~.
\nonumber
\end{equation}
To simplify notation we shall in the following write the continued fraction as
\begin{equation}
c_0(z)=\Delta_0/(z + \Delta_1/(z + \Delta_2/(z + ...)))~.
\label{7}
\end{equation}
The first $K$ of the continued fraction coefficients are uniquely determined 
by the corresponding first $K$ even frequency moments (\ref{mn}) 
through an iterative procedure described in Ref.~\onlinecite{Muller}. Of 
course, we just traded the short-time behavior of $C_0(t)$ for the large-$z$ 
behavior of $c_0(z)$, which does not bring us any closer to the desired 
$z \sim 0$ region. But, as described in detail in the book by Viswanath and 
M\"{u}ller, \cite{Muller} the relaxation function is uniquely determined by 
the sequence ${\Delta_k}$, which contains important information about the 
asymptotic behavior of $c_0(z)$. Namely, for the isotropic Heisenberg model 
(\ref{hamilton}), the $\Delta$-sequence grows with the index $k$ 
according to a power law, $\Delta_k \sim k^{\lambda}$, with 
$1 \leq \lambda \leq 2$, which fixes the high-frequency behavior of the
spectral density to $\Phi(\omega) \sim \exp{(-|\omega|^{2/\lambda})}$.
Moreover, oscillations of the odd (or even) continued fraction coefficients
around the $k^{\lambda}$ curve contain information on the {\it infrared} 
behavior of $\Phi(\omega)$. The simplest function that incorporates both the 
high- and low-frequency behavior typical of critical spin systems has the form
\begin{equation}
\bar{\Phi}(\omega)=\frac{2\pi}{\lambda \omega_0 \Gamma(\lambda(1 + \alpha)/2)}
\mid \frac{\omega}{\omega_0} \mid^{\alpha} 
\exp\left(- \mid \frac{\omega}{\omega_0} \mid^{2/\lambda}\right).
\label{phiform}
\end{equation}
The frequency moments of this function are known to be
\begin{equation}
\bar{M}_{2k}=\omega_0^{2k} \frac{\Gamma(\frac{\lambda}{2}(1 + \alpha + 2k))}
{\Gamma(\frac{\lambda}{2}(1 + \alpha))},
\label{phimoments}
\end{equation}
and the corresponding continued fraction coefficients $\bar{\Delta}_k$
can be calculated from them numerically.
Of course, an approximation of the spectral density $\Phi(\omega)$ of the 
system under study by the model density $\bar{\Phi}(\omega)$, with parameters
$\omega_0,\alpha,\lambda$ determined from a ``given'' sequence ${\Delta_k}$,
would be just marginally better than the often used gaussian ansatz.
Instead, M\"uller and collaborators (see Ref.~\onlinecite{Muller} and 
references therein) devised a more accurate procedure, which we describe 
here for completeness.

Suppose that we have calculated the first $K$ even moments of the true spectral
density $\Phi(\omega)$. Then we calculate the corresponding sequence
$\Delta_k$, and try to approximate it by the model sequence $\bar{\Delta}_k$
by minimizing the sum $\sum_{k=k_{min}}^{K} (\Delta_k - \bar{\Delta}_k)^2$
with respect to the parameters $\omega_0,\alpha,\lambda$. The lower cut-off
$k_{min}$ ($=3$ in our study) is necessary because the first few coefficients
$\Delta_0, \Delta_1, ...,\Delta_{k_{min}}$ tend to deviate significantly
from the asymptotic behavior represented by $\bar{\Phi}(\omega)$.
Having determined the parameters of the fit we may find exactly the
$K$-th level $terminator$ $\bar{\Gamma}_K(z)$ of the model relaxation 
function corresponding to $\bar{\Phi}(\omega)$
by (numerically) inverting the equation
\begin{equation}
\bar{c}_0(z)=\bar{\Delta}_0/(z +\bar{\Delta}_1/(z + ...+\bar{\Delta}_K/
(z + \bar{\Gamma}_K(z)))).
\label{10}
\end{equation}
The terminator thus incorporates information on the asymptotic behavior
of the $\bar{\Delta}_k$-sequence.

The relaxation function $c_0(z)$  in Eq.~(\ref{7})
is then approximated as
\begin{equation}
c_0(z)=\Delta_0/(z + \Delta_1/(z + ...\Delta_K/(z + \bar{\Gamma}_K(z)))),
\label{11}
\end{equation}
and thus, in addition to the correct large-$z$ behavior contained in the
first few exactly known continued fraction coefficients, through the terminator 
$\bar{\Gamma}_K(z)$, $c_0(z)$ also incorporates 
the correct small-$z$ behavior extracted from the
$\Delta_k$ sequence. Analytic continuation $z \rightarrow -i\omega^{\prime}$ 
then gives us the spectral function $\Phi(\omega^{\prime})$ in the whole
range of frequencies $\omega^{\prime}$. For the model spectral function
$\bar{\Phi}(\omega)$ of type (\ref{phiform}), such analytic continuation is
performed by hand and only requires numerical integration of well-behaved 
functions.

To study the spin dynamics at $finite$ temperature we have calculated
moments of $S_r(\omega)$ by the HTE technique. It
is well known that these
moments, $M_k$, can be expressed in terms of a thermal expectation value of
a $k$-fold commutator. \cite{Bohm,Muller} High temperature expansion can be
developed for these quantities by the cluster method. \cite{Baker} 
In fact, using the same set of clusters,
the expansions for the $k$-th moment will be complete to
order $\beta^{N-k}$, where $N$ depends on the size of the largest cluster
considered. We have calculated up to $N=22$ for all non-zero moments.
The equal-time correlation function is calculated to order $\beta^{20}$.
It is more convenient to do the calculations for the scaled
function $c_0(z)/M_0$, i.e. the one defined by the normalized
set of moments $\{1,M_2/M_0,...,M_{2K}/M_0\}$, and obtain the needed
function $c_0(z)$ by simple multiplication at the very end of
calculations. 

The behavior of the first 7 continued fraction coefficients of the spin 
autocorrelation function $\langle S^z_0(t)^zS_0(0)\rangle$
as a function of inverse temperature is shown in Fig.~1. From the
relation between $M_{2k}$ and the $\Delta_k$ sequences \cite{Muller}
we have $\Delta_0=M_0$, and hence $\Delta_0=1/4$ irrespective of
temperature in that case. Another property of the $\Delta_k$'s that make them 
convenient for numerical computations is that they are all numbers 
of order 1, whereas the corresponding frequency moments grow very fast with
the index, for example $M_{14}(T=\infty)=166988876$. 
Continued fraction coefficients with higher index $k$ are related to the
moments $M_{2k}$ which are determined by spin correlations at
larger distances, and hence are more temperature sensitive.
The HTE ceases to work for $\Delta_7$ at
$\beta \sim 0.75$ where it starts to change rapidly, and for $\Delta_6$
at $\beta \sim 1.5$. Since we want to reach the lowest possible
temperature, we restrict ourselves to first $6$ ($k=0...5$)
coefficients of the sequence, which permits analysis up to $\beta \sim 2$.
Of course, the shorter the $\Delta_k$ sequence the more uncertain becomes
the determination of parameters $\omega_0,\alpha,\lambda$, and in each
particular study a try-and-see approach should be used to find a
compromise between these two conflicting requirements. We found that
at $\beta \leq 0.5$ results obtained with $K=7$ and $K=5$ do not
differ much, and upto $\beta=2$ the sequence $\Delta_0 ... \Delta_5$ 
is stable and reliable.

In Fig.~1 we also show the $k$-dependence of the $\Delta_k$
sequence for $S_R(\omega)$ (Eq.(\ref{t11})) for $R=0$ and $R=-0.5$. The latter
has a vanishing form factor at $k=0$ and thus has no contributions from the 
diffusive modes. It is evident that the former sequence exhibits an odd-even
oscillation, suggesting an infrared singularity, but this
is absent from the latter sequence. This ability to recognize
the presence or absence of diffusive modes at such a simple
level shows the power of the recursion method.

The temperature dependence of the parameters of the fit, Eq.(\ref{phiform}),
is shown in Fig.~\ref{fit}. Notice the drastic difference in  $\alpha$, 
the power of the infrared singularity for $R=0$ and $R=-0.5$:
in the latter case it is always zero, whereas in the former it decreases 
rapidly with temperature, demonstrating the suppression of the diffusive 
behavior at low $T$.

The overall quality of the described procedure can be estimated
by direct term-to-term comparison of the real sequence $\Delta_k$ with
the model one $\bar{\Delta}_k$, given in Fig.\ref{delta} for $k=3,4,5$.
Deviations of $\bar{\Delta}_k$ from $\Delta_k$ are most
pronounced at high temperatures for the $R=0$ case, where the
low-frequency infrared divergence is the strongest, and almost
disappear for $T$ below $J$. The deviations at high $T$ can be
reduced by working with longer $\Delta_k$-sequences, as was mentioned
above, but this would significantly reduce the range of applicability
of our calculations. The absolute value of the discrepancy between 
$\Delta$'s does not exceed $5\%$ in the worst case. 

\subsection{Stochastic Series Expansions and the Maximum Entropy Method}

The stochastic series expansion QMC method \cite{Sandvik1,Elescorial} is an 
improved variant of the so called Handscomb's technique.\cite{Handscomb} It 
is based on importance sampling of terms of the power series expansion of 
exp$(-\beta\hat H)$, which for a finite system at finite $\beta$ can be 
carried out to all important orders, without introducing systematic errors.
The current formulation of the method is described in 
Ref.~\onlinecite{Elescorial}, and has previously been applied to both 
static and dynamic properties of Heisenberg models in several different 
geometries.\cite{SSEwork,Sandvik2} Here we briefly review how imaginary-time 
independent correlation functions are evaluated using this technique.

The Hamiltonian for an $N$-site periodic chain is first written as
\begin{equation}
\hat H = -{J\over 2} \sum\limits_{b=1}^{N} 
\bigl (\hat H_{1,b} - \hat H_{2,b} \bigr ) + {NJ\over 4},
\label{ham2}
\end{equation}
where the operators $\hat H_{1,b}$ and $\hat H_{2,b}$ are defined as
\begin{mathletters}
\begin{eqnarray}
\hat H_{1,b} & = & 2 \bigl ( \hbox{$1\over 4$} - S^z_{b}S^z_{b+1} \bigr ), \\
\hat H_{2,b} & = & S^+_{b}S^-_{b+1} + S^-_{b}S^+_{b+1} .
\end{eqnarray}
\label{hops}
\end{mathletters}
The partition function, $Z={\rm Tr}\lbrace {\rm e}^{-\beta\hat H} \rbrace$, 
is transformed into a sum suitable for importance sampling techniques by
expanding ${\rm e}^{-\beta\hat H}$ in a power series, and writing the
trace as a sum over diagonal matrix elements in the standard basis
$\lbrace |\alpha \rangle \rbrace= \lbrace |S^z_1,\ldots,S^z_N \rangle 
\rbrace$. This gives
\begin{equation}
Z = \sum\limits_\alpha \sum\limits_n \sum\limits_{S_n} {(-1)^{n_2} \over n!}
\Bigl ({\beta \over 2} \Bigr )^n \Bigl \langle \alpha \Bigl | 
\prod\limits_{l=1}^n \hat H_{a_l,b_l} \Bigr | \alpha \Bigr \rangle ,
\label{partition}
\end{equation}
where $S_n$ denotes a sequence of index pairs defining the operator string 
$\prod_{l=1}^n \hat H_{a_l,b_l}$,
\begin{equation}
S_n = [a_1,b_1][a_2,b_2]\ldots [a_n,b_n],\quad
a_i \in \lbrace 1,2\rbrace$, $b_i \in \lbrace 1,\ldots ,N_b \rbrace,
\label{sn}
\end{equation}
and $n_2$ denotes the total number of index pairs (operators) $[a_i,b_i]$ 
with $a_i = 2$. For even $N$ (or any $N$ for an open chain) $n_2$ is 
even for all non-zero terms in Eq.~(\ref{partition}). All terms 
are then positive, and can be stochastically sampled using standard 
Monte Carlo techniques in the space of index sequences and basis states.

The simulation is carried out using two different elementary modifications 
of the index sequence, taking into account the constraints imposed by the 
fact that the operators defined in Eqs.~(\ref{hops}) are allowed to operate 
only on antiferromagnetically aligned spin pairs, and that the operator 
product corresponding to $S_n$ must propagate the state $|\alpha \rangle$ 
onto itself. The power $n$ is changed by inserting or removing single diagonal
operators $[1,b]$, and the number of spin flipping operators is changed by 
pair-substitutions $[1,b],[1,b] \leftrightarrow [2,b],[2,b]$ (the two 
operators selected for updating are typically not adjacent in the sequence). 
The grand canonical ensemble with fluctuating total magnetization, 
$m^z=\sum_i S^z_i$, can be studied for $T/J \agt 0.08$ by also performing 
spin flips in the states. At lower temperatures the acceptance rate for 
such flips becomes very low, and it is then more convenient to study the 
canonical ensemble with $m^z=0$. 

QMC calculations can access the dynamic structure factor only through the 
corresponding correlation function in imaginary time,
\begin{equation}
C_{r_1,r_2} (\tau)= \langle S^z_{r_1} (\tau) S^z_{r_2} (0) \rangle .
\label{taucorr}
\end{equation}
In the stochastic series expansion method, such a correlation function is
estimated by measuring the correlations between states 
$| \alpha (p) \rangle = |S^z_1[p],\ldots,S^z_N[p] \rangle$
obtained by propagating
$| \alpha \rangle$ in Eq.~(\ref{partition}) with $p$ of the operators
in the product:
\begin{equation}
| \alpha (p) \rangle = \prod\limits_{l=1}^p \hat H_{a_l,b_l} |\alpha \rangle ,
\quad | \alpha (0) \rangle = | \alpha \rangle .
\label{propagated}
\end{equation}
The expression for the imaginary-time dependent spin-spin correlation
function is \cite{Sandvik1}
\begin{equation}
C_{r_1,r_2}(\tau) =
\biggl\langle \sum\limits_{m=0}^n  {\tau^m (\beta -\tau )^{n-m}n!\over 
\beta ^n (n-m)!m!}
\bar C_{r_1,r_2}(m) \biggr\rangle ,
\label{taudia}
\end{equation}
where 
\begin{equation}
\bar C_{r_1,r_2}(m) =
{1\over n+1} \sum\limits_{p=0}^{n} S^z_{r_1}[p]S^z_{r_2}[p+m] .
\end{equation}
The periodicity of the propagated states imply that $S^z_r[p+n]=S^z_r[p]$.
Note that any value of $\tau$ is accessible, in contrast to worldline
methods where $\tau$ must be an integer multiple of the time-slice width
used in the simulation.\cite{worldline} The corresponding static 
susceptibility $\chi_{r_1,r_2}$ (i.e., the real-space version of 
Eq.~(\ref{Kubo})), can  be directly obtained by integrating over 
$\tau$ in (\ref{taudia}). The result is \cite{Sandvik1}
\begin{equation}
\chi_{r_1,r_2}= \Bigl \langle {\beta\over n(n+1)} 
\Bigr ( \sum\limits_{p=0}^{n-1} S^z_{r_1}[p] \Bigr )
\Bigr ( \sum\limits_{p=0}^{n-1} S^z_{r_2}[p] \Bigr ) 
+ \beta {\bar C_{r_1,r_2}(0)\over n+1} \Bigr\rangle .
\label{diasus}
\end{equation}

The relation between $C_{r_1,r_2}(\tau)$ and the dynamic structure factor 
defined in Eq.~(\ref{sqwdef}) is
\begin{equation}
S_k (\tau)= {1\over \pi} \int\limits_0^\infty 
d\omega S(k,\omega)K(\omega,\tau),
\label{contint}
\end{equation}
where $S_k (\tau)$ is the Fourier transform
\begin{equation}
S_k (\tau)= {1\over N}\sum\limits_{r_1,r_2} {\rm e}^{-ik(r_2-r_1)}
C_{r_1,r_2}(\tau),
\label{taucorrq}
\end{equation}
the kernel is
\begin{equation}
K(\omega,\tau) = {\rm e}^{-\tau\omega} + {\rm e}^{-(\beta-\tau)\omega},
\end{equation}
and $S(k,-\omega) = {\rm e}^{-\beta\omega}S(k,\omega)$. The analytic 
continuation of the QMC data for $S_k(\tau)$, i.e., inversion of 
Eq.~(\ref{contint}), is carried out using the Max-Ent method,
\cite{maxent1,maxent2} which we very briefly review here for 
completeness.

For a given momentum transfer $k$, $S(k,\omega)$ is paremetrized as a sum 
of $\delta$-functions at frequencies $\omega_n$, $n=1,\ldots N_\omega$,
\begin{equation}
S(k,\omega) = 
\sum\limits_{n=1}^{N_\omega} S_n \delta(\omega-\omega_n).
\label{ssum}
\end{equation}
The ``classic'' variant of the Max-Ent method used here amounts to 
determining the coefficients $S_n$ that minimize the quantity
\begin{equation}
Q = \chi^2/2 - \alpha E.
\label{qdef}
\end{equation}
Here $\chi^2$ is the deviation of the imaginary-time function 
$\bar S_k(\tau)$ corresponding to a particular set of spectral weights 
$\lbrace S_1,\ldots ,S_n \rbrace$ in Eq.~(\ref{ssum}) from the QMC estimate
$S_k(\tau)$ and its statistical error $\sigma (\tau)$, which is available
for a discrete set of times $\tau_1,\ldots ,\tau_{N_\tau}$: \cite{cnote}
\begin{equation}
\chi^2 = \sum\limits_{i=1}^{N_\tau} [S_q(\tau_i) - \bar S_q(\tau_i)]^2 
\bigr /\sigma (\tau_i)^2.
\label{chi2}
\end{equation}
$E$ is the entropy, defined with respect to a ``default model'' $m(\omega)$. 
Using a default which is constant for $\omega > 0$ and satisfies 
$m(-\omega)={\rm e}^{-\beta\omega}m(\omega)$, the entropy is (with 
$S(k,\omega)$ assumed normalized to unity), 
\begin{equation}
E = - \sum\limits_{i=1}^{N_\omega} S_n \ln(S_n) K(\omega_n,0).
\end{equation}
The parameter $\alpha$ in Eq.~(\ref{qdef}) is determined iteratively using 
a criterion derived using Bayesian logic, leading to the most probable 
$S(k,\omega)$ compatible with the QMC data and its errors, as discussed in 
detail in Ref.~\onlinecite{maxent2}. Typically, on the order of 
$N_\omega=100-200$ frequencies are used in Eq.~(\ref{ssum}). The amplitudes 
$S_n$ then form a smooth curve representing the frequency dependence of
$S(k,\omega)$.

\subsection{Comparisons with Exact Diagonalization}

Accuracies of calculations of dynamic quantities using the HTE and QMC 
methods are difficult to assess rigorously. Comparing
results obtained in the two different ways provides for a good test. However,
the results will never agree completely (In contrast, for static quantities
the results agree perfectly in the regime where HTE performs well, as
discussed in Sec.~III-C.), and it is therefore useful to check the results 
against other calculations as well. 

For a small system, all the eigenstates can be obtained exactly by numerically 
diagonalizing the Hamiltonian. Using the translational invariance and the
conservation of the $z$-component of the spin, the Hamiltonian consists
of blocks corresponding to all the combinations of the magnetization $m^z$
and the momentum $k$. For a 16-site system the largest blocks have 810 
states, and can easily be diagonalized on a workstation. The next two
appropriate sizes, $N=18$ and $N=20$, have largest-block sizes of 2704 and 
9252 states, respectively, and could be studied with some more effort.
Here we consider only $N=16$.

The exact dynamic structure factor of a finite system is a set of 
$\delta$-functions with positions given by the energies of the excited
states, and amplitudes given by the squares of the corresponding matrix 
elements of the operator $S^z_k$. For a 16-site system, the number of 
contributing $\delta$-functions is very small at $T=0$, and it is 
not possible to carry out a meaningful comparison with the other methods. 
As the temperature is raised, the number of $\delta$-functions with 
significant weights increases rapidly, and a relatively smooth spectrum can be 
obtained by using some reasonable broadening of the individual 
$\delta$-functions. The results can of course not be expected to completely 
represent the thermodynamic limit, but at temperatures where the correlation 
length is much smaller than the system size meaningful comparisons should be 
possible. We here compare HTE, QMC, and exact diagonalization results 
for $S(k,\omega)$ at $k=\pi/2$.

Fig.~\ref{comp} shows results of all the methods at temperatures $T/J=1.0$ 
and $0.5$. We have chosen to represent the exact results as histograms with 
a bin width $\Delta_\omega=0.1J$. The $n$th bin contains the integrated 
spectral weight in the frequency range $[(n-1)\Delta_\omega,n\Delta_\omega]$. 
On this frequency scale, the results still have structure due to the finite 
size, but nevertheless exhibit over-all shapes that the other results 
can be compared with. Indeed, the HTE spectra have shapes that very 
well match the histograms. The QMC + Max-Ent results are somewhat broader
and have more rounded shapes (less asymmetric), as would also be expected, 
but still represent quite reasonable distributions of spectral weight. 
The QMC results were calculated for a system of $N=128$ sites. Results 
obtained for the same size, $N=16$, as the exact diagonalization, look 
very similar, which also indicates that finite size effects are small at 
these temperatures (for the momentum considered here). The relative 
statistical errors of the imaginary-time data used were in the range 
$10^{-4} - 10^{-3}$, which is typical for all the QMC results discussed 
in this paper (the absolute error is typically in the fifth decimal
place of the result).

The limit $\omega \to 0$ is of special interest, as it determines the
spin-lattice relaxation rate. The results shown in Fig.~\ref{comp} indicate
that meaningful results can be obtained for this quantity. The differences 
between the HTE and QMC + Max-Ent results are typically 10-20\%. We cannot 
rigorously establish which calculation is more accurate in the $\omega \to 0$ 
limit, but based on the better agreement with the over-all shape between 
HTE and exact diagonalization in the whole frequency range, we expect HTE 
to be more accurate in the temperature regime where it performs well ($T/J 
\agt 0.5$). At lower temperatures only QMC + Max-Ent results are available, 
since exact diagonalization also does not provide accurate results below 
$T/J \alt 0.5$, especially for $\omega \to 0$. Based on high-temperature
comparisons such as those presented here, we expect the error of the
QMC + Max-Ent calculations at lower temperatures to be of the order
10-20\%.

\section{Results}

The HTE method is best suited for studying the dynamics of $k$-integrated 
quantities. Apart from the results for $k=\pi/2$ presented above, we 
therefore discuss HTE results 
mainly for momentum averages and the spin-lattice relaxation rate $1/T_1$. 
With the QMC method we have calculated the imaginary-time correlation functions
needed to obtain the full $S(k,\omega)$ for systems with up to $N=128$, down 
to temperatures $T/J=1/8$. Results for $1/T_1$, as well as the transverse rate
$1/T_{2G}$, were already presented in Ref.~\onlinecite{Sandvik2}, where 
$1/T_1$ was calculated for slightly larger systems by analytically continuing 
weighted imaginary-time correlation averages for space separations $r\le 2$, 
corresponding to the model form factor discussed in Sec.~II. Here 
$S(k,\omega)$ is first calculated for all momenta, and the momentum averaging 
is carried out after the analytic continuations. This method, though much more
cumbersome and time consuming, should be more reliable for studying long-time 
tails such as those resulting from spin diffusion at high temperatures.

Below, in Sec.~III-A, we first consider various aspects of the frequency
and momentum dependence of the dynamic structure factor. In III-B we discuss 
the spin-lattice relaxation rate, and recent experimental results for 
Sr$_{\rm 2}$CuO$_{\rm 3}$ and CuGeO$_{\rm 3}$. In III-C we discuss
the scaling behavior of the dynamic susceptibility, and how it is affected
by logarithmic corrections. We present explicit calculations of the
logarithmic corrections to the temperature dependence of the staggered
structure factor and the staggered susceptibility.

\subsection{The Dynamic Structure Factor}

Results for $S_0(\omega)$ (corresponding to a form factor $A(k)=1$), 
obtained  using the HTE technique at different temperatures are shown in 
Fig.~\ref{swhte}. As discussed earlier, we expect these results to be reliable
down to $T/J=0.5$. We find that at very high temperatures there is strong 
$\omega^{-\alpha}$ divergence which diminishes as $T$ is decreased, and 
spectral weight is transferred to a broad peak at $\omega \sim 1.5$. At $T=1$, 
a peak at $\omega \sim 1.5$ is evident together with an infrared peak, 
which is still strong at this temperature. However, at $T=0.5$ the 
diffusion peak is almost invisible and most of the spectral weight is 
concentrated at higher frequencies.

$S_0(\omega)$  calculated using QMC and Max-Ent analytic continuation 
is shown in Fig.~\ref{swqmc}, for temperatures down to $T/J=1/8$. Results for 
systems with $N=64$ and $N=128$ are compared, in order to assess effects of 
the system size. The agreement with the HTE results is quite good for 
$T/J=1.0$ and $0.5$. The diffusive $\omega \to 0$ peak at $T=1.0$ is somewhat 
higher and narrower in the HTE result, whereas it is somewhat more pronounced
in the QMC results at $T/J=0.5$. At high temperatures, the $\omega \to 0$
peak height grows with the system size in a QMC calculation, since the 
diffusive contributions are cut-off at the momentum $k_1 = 2\pi/N$ in a 
finite system. Considering the intrinsic difficulties of numerical analytic 
continuation, in particular of QMC data, the agreement between the HTE and 
QMC + Max-Ent results has to be considered quite satisfactory.

Apart from the $\omega \approx 0$ peak at high temperatures, the differences 
between the $N=64$ and $N=128$ results are likely mainly due to statistical 
fluctuations in the imaginary-time data, which are amplified in the real 
frequency spectra. The dominant peak at $\omega/J \approx 1.5$ is very 
pronounced at $\beta=8$, and does not exhibit 
much size dependence. The position of the peak can be understood on the 
basis of the $T=0$ two spinon continuum. The Bethe ansatz solution gives an 
exact expression for the lower edge,\cite{Cloiseaux} $\omega_{\rm min} (k) 
= (\pi/2)J\sin{(k)}$. The upper bound is known to be
\cite{Muller2} $\omega_{\rm max}(k)=\pi J\sin{(k/2)}$. 
Since the dominant spectral weight is concentrated at the lower edge, one 
can expect a maximum in the momentum average $S_0(\omega)$ at $\omega=\pi/2$, 
arising from momenta $q \approx \pi/2$, where the lower bound has the smallest 
$q$-dependence. This is indeed the position of the maximum seen in 
Fig.~\ref{swqmc} at $\beta=8$. 

A maximum at lower frequency also develops in $S_0(\omega)$ for 
$T < 0.25J$. It is due the gradual change from relaxational to propagating 
behavior for modes with momenta $3\pi/4 \alt k \le \pi$. The peak sharpens 
and moves towards $\omega =0$ as the temperature is lowered (this trend is
seen also in results for lower temperatures, not shown here). This maximum 
is expected from the scaling form for the dynamic susceptibility first 
derived by Schulz \cite{Schulz,Sachdev} ($\chi^{\prime \prime}(\omega) 
\sim {\rm tanh}(\omega/2T)$ for $T \ll J$), and has also been discussed 
in the context of neutron scattering experiments on
Cu(C$_6$D$_5$COO)$_2$.3D$_2$O.\cite{Aeppli}

Fig.~\ref{s0qmc} shows the momentum dependence of $S(k, \omega \to 0)$ at 
several temperatures. It is clear that the low-frequency weight is strongly 
peaked near $k=0$ at $T/J=1$, but shifts to $k=\pi$ as the temperature is 
lowered. The strong increase as $k \to 0$ at high temperatures is clearly 
not captured completely in a small system, due to the discreteness of the 
momentum. Apart from this long-wavelength cut-off, there is no size dependence
between $N=64$ and $128$ within the fluctuations of the data. We have not 
explicitly calculated statistical errors of the results, but one can get an 
impression of their order from the (rather low) degree of non-smoothness of 
the curves. As discussed in Sec.~II-C, there may  be some systematical errors
present as well, due to the unavoidable bias of the Max-Ent method used for 
the analytic continuation. In particular, this may be the case close to both 
$k=0$ and $k= \pi$, in the neighborhood of the points where low frequency 
weight first starts to appear (i.e. at temperatures where there is an 
intermediate momentum regime with vanishing low-frequency weight). In these 
regimes $S(k, \omega \to 0)$ may be over-estimated due to broadening effects,
i.e., low frequency weight may be seen in Fig.~\ref{swqmc} where in fact 
the actual modes only begin to have significant weight slightly above 
$\omega =0$. Away from these regimes, we expect systematic errors of at 
most 10-20\%, as discussed in Sec.~II-C.

We now discuss the full dynamic structure factor in the temperature regime 
where there is the most significant shift in spectral weight from $k \approx 
0$ to $k\approx \pi$. We present $N=128$ results for $T/J=1.0$, $0.5$, 
and $0.25$ graphed in two different ways. First, in Fig.~\ref{sqw3d}, we show 
3D graphs with curves of $S(k,\omega)$ for each individual $k$. This clearly 
demonstrates how the narrow $k \approx 0$ peak present at $T/J=1$ is 
significantly reduced at $T/J=0.5$, where there is also a massive build-up 
of spectral weight at momenta close to $k=\pi$. The maximum at $k=\pi$ is 
not yet very pronounced at this temperature, however. At $T/J=0.25$ the 
$k=0$ peak has vanished almost completely, and the distribution of spectral 
weight starts to resemble what would be expected from the $T=0$ two-spinon 
continuum. Again, the lack of smoothness along momentum cross-sections gives an 
impression of the considerable amplification by analytic continuation of the 
very small statistical errors in the imaginary-time data. The concentration 
of the weight between the lower and upper bounds of the two-spinon spectrum 
is seen more clearly in the plots of Fig.~\ref{sqw2d}. Here the intensity 
is represented by shades of gray in the $(k,\omega)$-plane, and the $T=0$ 
bounds are also shown. It is clear that there is very little weight above 
the upper bounds even at high temperatures, whereas there is significant
weight below the lower bound.

One may still wonder how well the Max-Ent method captures the true temperature
dependence of $S(k,\omega)$. In the previous low-temperature calculations
by Deisz {\it et al.},\cite{Deisz2} considerable weight was observed below 
the rigorous lower bound even at temperatures as low as $T/J = 1/24$, and 
the expected concentration of weight at the lower edge was not well 
reproduced. Above we have shown that our high-temperature results agree 
well with HTE calculations. In order to further investigate the broadening 
effects due to Max-Ent analytic continuation, we have also carried out 
simulations for the system size and temperature considered by Deisz {\it et 
al.} ($N=64$, $\beta=24$). Fig.~\ref{s34pi} shows our result for $k=3\pi/4$, 
which can be compared with Fig.~9 of Ref.~\onlinecite{Deisz2}. Our result
indeed peaks at the lower bound, and is significantly less broadened towards 
lower frequencies. This probably reflects a higher accuracy in the underlying 
imaginary-time data. The broadening that is present at $\beta=24$ may still 
be partly due to temperature effects, but is likely mainly Max-Ent induced. 
This kind of broadening should be a problem primarily in cases where the
lower edge is sharp, i.e., at very low temperatures. It will be present
to some extent also for temperatures and momenta where there is very 
little low-frequency weight, and, as already discussed above, may then lead 
to an over-estimation of $S(k,\omega \to 0)$. The broadening below the lowest 
bound seen in the results of Figs.~\ref{sqw3d} and \ref{sqw2d} is considerably
larger than in Fig.~\ref{s34pi}, and we therefore believe that it is mainly 
due to real temperature effects, with only minor distortions due to Max-Ent 
bias.

\subsection{Spin-Lattice Relaxation Rate}

Next we discuss our calculations of the nuclear spin-lattice relaxation 
rates $1/T_1$, for different hyperfine couplings parametrized by the 
ratio $R$ in Eq.~(\ref{t1}). 

The QMC results are obtained by averaging the low-frequency dynamic 
structure factor shown in Fig.~\ref{s0qmc} (for $N=128$). As already 
mentioned, this method differs from the approach previously used in 
Ref.~\onlinecite{Sandvik2}, where the averaging was done for the 
short-distance imaginary-time correlation functions, and the analytic 
continuation carried out subsequently. With exact imaginary-time data, the 
two approaches would yield identical results, but in the presence of 
statistical errors the Max-Ent method will bias the outcome differently in
the two approaches. As an illustration of this, results for the on-site 
dynamic structure factor $S_0 (\omega)$ obtained using the two different 
orders of averaging and analytic continuation are shown in Fig.~\ref{comps0}. 
It is clear that the analytic continuation of the on-site function $S_0(\tau)$
misses much of the diffusive behavior for $\omega \to 0$ at the temperatures 
where these contributions are the most important. Based on the previous 
comparisons with HTE results at high temperatures, we believe that the
approach of continuing $S(k,\omega)$ individually for each $k$ before 
averaging is more accurate. This can also be understood on grounds that 
the frequency dependence of $S(k,\omega)$, for a given $k$, has less 
structure than the momentum average, and is therefore easier to reproduce
with a high-entropy curve (which is favored by the Max-Ent method).

The HTE + recursion method is applied to the short-distance correlations, 
according to Eq.~(\ref{t11}). The zero frequency limit will produce a 
divergent $1/T_1$ when infrared singularities are present. However, 
the very low-frequency, long-wavelength limit of our results may not be 
accurate and should be viewed with some caution as the method is based on
a short-time expansion, involving only finite spin clusters.
We have therefore chosen not to focus on the exact form of the $\omega \to 0$,
$q \to 0$ behavior. For calculating $1/T_1$ we set the 
nuclear resonance frequency to $0.01J$. In a calculation using QMC + Max-Ent, 
the divergence is cut off due to the finite size, and may also be rounded due 
to resolution effects. In real materials, several energy scales including 
anisotropy and coupling between chains serve to determine the cutoff frequency.

Results for $1/T_1$ from the recursion method are shown for several different 
values of $R$ in Fig.~\ref{t1fig}, along with the QMC + Max-Ent results for 
$R=0$ and $R=-0.5$. The agreement between the HTE and QMC calculations is 
quite good for $T/J \ge 0.5$, and gives us confidence in the QMC data at 
lower temperatures. The main difference from the previous \cite{Sandvik2}
QMC + Max-Ent results is that the diffusive contributions for $R =0$ at 
$T \agt 0.5$ are stronger, as already discussed. At low temperatures the 
present results are $\approx$20 percent lower than the previous results, both 
for $R=0$ and $R=-0.5$. It is likely that the present results at low 
temperatures are still somewhat too high (likely $10-20$ percent), due 
to the broadening effect of the max-ent method discussed above (see also 
Ref.~\onlinecite{SSS}, for a comparison of results for the ratio 
$T_{2G}/\sqrt{T}T_1$).

Sachdev calculated $1/T_1$ using the scaling form of $S(k,\omega)$ obtained 
from bosonization and conformal field theory.\cite{Sachdev,Sachdev2,Schulz}
This gives a temperature independent rate, which however is 
expected to be modified by logarithmic corrections to yield a logarithmic 
divergence as $T\to 0$.\cite{Sachdev} This is clearly in qualitative agreement
with the above results for $R=-0.5$ (where the diffusive contributions 
neglected in the theory are filtered out), but the accuracy is not high 
enough to extract a form for the low-temperature behavior. We have previously 
discussed other effects of logarithmic correction on the NMR relaxation 
rates.\cite{SSS} In particular, it can be noted that the ratio 
$T_{2G}/\sqrt{T}T_1$ is modified from the universal temperature-independent 
form, approaching the universal value with infinite slope as $T \to 0$ 
(in a manner similar to the bulk susceptibility, as discussed  
in \cite{bethe1}).

We now briefly discuss recent experimental results. Takigawa {\it et al.}
measured both $1/T_1$ and $1/T_{2G}$ in Sr$_{\rm 2}$CuO$_{\rm 3}$,
\cite{Takigawa} for which an exchange $J \approx 2000$K had been previously 
deduced from susceptibility measurements.\cite{Srchain} The hyperfine form 
factor could be accurately extracted \cite{Takigawa2} using an impurity 
effect on the NMR line shape predicted by Eggert and Affleck.\cite{Eggert} 
Hence, there are no free parameters, and non-ambiguous comparisons with model
calculations can be carried out. Takigawa {\it et al.}~concluded that
the agreement with the $1/T_{2G}$ calculation of Ref.~\onlinecite{Sandvik2}
was good. For $1/T_1$ the lowest temperature studied in 
Ref.~\onlinecite{Sandvik2} corresponds to $\approx 300$K, which was the 
highest temperature studied experimentally. At this temperature, the 
previous QMC + Max-Ent result was $\approx 40 \%$ higher than the experimental
value. As noted above, our improved method of calculating $1/T_1$ gives a 
value $\approx 20\%$ lower than before, and hence this discrepancy is now 
largely reconciled (the remaining differences can likely be explained by
uncertainties in the experimental values of $J$ and the hyperfine couplings,
and by remaining errors in the numerical result). As discussed above 
and in Ref.~\onlinecite{SSS}, also the slight temperature dependence observed 
for $T_{2G}/\sqrt{T}T_1$ can be explained theoretically, and is largely due to 
logarithmic corrections. Hence, all the analytical and numerical results are 
now in good agreement with the experiments, indicating that 
Sr$_{\rm 2}$CuO$_{\rm 3}$ indeed is an almost perfect realization of the 
spin-half Heisenberg chain with only nearest-neighbor interactions. 

We also note that some contributions from diffusive processes, signaled by a 
dependence of $1/T_1$ on $\omega_N$ (i.e., the external field strength) were
explicitly observed in Sr$_{\rm 2}$CuO$_{\rm 3}$.\cite{Takigawa} It
would clearly be interesting to study this material also at higher 
temperatures, where our numerical results indicate that a considerably 
stronger diffusive behavior should be observed.

Another quasi-1D compound which has been studied very actively recently
is CuGeO$_{\rm 3}$, which undergoes a spin-Peierls (SP) transition at
$T \approx 14$K.\cite{CuGeO} This material has a stronger coupling between 
the chains than Sr$_{\rm 2}$CuO$_{\rm 3}$, and also is expected to have a 
non-negligible next-nearest-neighbor interaction $J_2$.\cite{J2papers} 
$1/T_1$ exhibits a strong, almost linear, decrease with decreasing 
temperature, with a reduction by almost a factor of $5$ between $T\approx J$ 
and $T \approx J/5$ (If $J_2$ indeed is significant, $J$ should here be
considered an effective coupling constant).\cite{Itoh,Horvatic} Itoh {\it et 
al.} argued that the transferred hyperfine term is small,\cite{Itoh} and hence 
$R \approx 0$ in our notation. Our results shown in Fig.~\ref{t1fig} then 
indicate a reduction by a factor of 2 in the above temperature regime, 
which is significantly different from the experiment. On the other
hand, Fagot-Revurat {\it et al.} \cite{Horvatic} argued that there is a 
significant transferred hyperfine coupling, and they were able to obtain a 
reasonable fit to the previous $1/T_1$ results for
$R \approx 0.25$.\cite{Sandvik2} Notice that 
this value of $R$ leads to a strong enhancement of contributions from 
diffusive, $k \approx 0$, modes, and as a result, a stronger temperature 
dependence of $1/T_1$ (as compared with, e.g., the $R=0$ case). Numerical 
calculations for the NMR rates including $J_2$, inter-chain couplings, 
and dynamic phonons have yet to be carried out, and would clearly be very 
useful for determining the importance of additional interactions beyond 
the standard Heisenberg chain in a realistic model of CuGeO$_{\rm 3}$. The 
fact that the temperature dependence of $1/T_1$ above the SP transition 
extrapolates to a positive value at $T=0$ \cite{Itoh,Horvatic} indicates 
that the frustrating $J_2$ coupling by itself is not sufficient for opening 
a gap, but it may clearly contribute to stabilizing the dimerized state.

\subsection{Scaling Behavior}

In this section we discuss the low-temperature scaling behavior of the
staggered susceptibility. At a $T=0$ quantum-critical point with $z=1$, 
conventional quantum critical scaling implies that the dynamic staggered 
susceptibility has the scaling form \cite{CSY}
\be
\chi(q,\omega)={a\over T^{2-\eta}}\phi\big({cq\over k_BT},{\omega\over k_BT}
\big),
\ee
where $a$ is a non-universal number and $\phi(x,y)$ is a universal complex 
function of both arguments and $q$ measures deviations from the 
antiferromagnetic wave-vector; $q=\pi-k$. This in turn also implies that
near the antiferromagnetic wave-vector, the equal-time structure
factor $S(q)$ and the antiferromagnetic susceptibility, $\chi(q)$
have scaling forms
\begin{mathletters}
\begin{eqnarray}
S(q)/S(0) & = & f_1(cq/T), \\ 
\chi(q)/\chi(0) & = & f_2(cq/T).
\end{eqnarray}
\end{mathletters}
However, for the Heisenberg chain these scaling relations are violated by
logarithmic factors, which are produced by marginally irrelevant operators
describing interaction between left- and right-moving currents \cite{bethe1}.
The failure of scaling for $\chi(q)$ was discussed earlier, and a way to 
take the logarithmic corrections into account analytically
was proposed.\cite{SSS} Here, in Fig.~\ref{sqfig}, we show the scaling plot 
for $S(q)$. Substantial systematic deviations from scaling in $q/T$ are 
apparent even at the lowest temperatures accessible to us. We also show
results for the quantity $S'(q)$, obtained by subtracting the ferromagnetic 
(uniform) component of the spin-spin correlation function from the numerical 
data for $S_r(0)$ before Fourier transforming to momentum space. The uniform 
term is given by $-[T/(2c{\rm sinh}(\pi T r/c)]^{-2}$, and contains no 
adjustable parameters.\cite{Sachdev} $S'(q)$ then contains fluctuations with 
momenta around the antiferromagnetic wave-number $\pi$ only, i.e. with 
$q \sim 0$. However, $S'(q)/S'(0)$ also does not exhibit scaling in $q/T$, 
demonstrating the importance of subleading non-universal corrections. Unlike 
what is the case for the static susceptibility,\cite{SSS} which is dominated
by low-frequency fluctuations, these deviations cannot be explained by 
taking only long-distance logarithmic corrections into account.

Turning now to scaling in $\omega /T$, we discuss the scaling amplitudes at 
$q=0$, i.e. the staggered structure factor and the staggered susceptibility.
Including the leading logarithmic corrections, the sum rules (\ref{sumrules}) 
give the low-temperature forms
\begin{mathletters}
\begin{eqnarray}
S(q=0) & = & D_s [\ln{(T_s/T)}]^{3/2}, \\
\chi(q=0) & = & D_\chi T^{-1} [\ln{(T_\chi/T)}]^{1/2}.
\end{eqnarray}
\label{s0x0}
\end{mathletters}
In Fig.~\ref{sxpi} these quantities are graphed in such a way that the 
behavior vs.~$\ln{(J/T)}$ should be linear if the above asymptotic forms
hold. The HTE results shown are from several different differential 
approximants. The agreement between the HTE and the QMC data is quite good 
down to $\beta\approx 6-8$, where the HTE approximants start to deviate 
from each other and from the QMC results, indicating that the HTE method 
becomes unreliable. Linear behavior consistent with Eqs.~(\ref{s0x0}) is 
indeed observed in the QMC data for $J/T \agt 4$. Fitting lines to the 
results in this regime gives $D_s=0.094 \pm 0.001$, $D_\chi=0.32 \pm 
0.01$, $T_s=18.3 \pm 0.5$, and $T_\chi= 5.9 \pm 0.2$. 

The difference in $T_s$ and $T_\chi$ may be due to divergent contributions 
to $S(0)$ from short distances, careful treatment of which would 
require a separate cutoff. Another important reason is the ferromagnetic 
contribution to the spin-spin correlation function, already discussed above. 
This contributes to both $S(0)$ and $\chi (0)$. In Fig.~\ref{sxpi} we 
also show results for the quantities $S'(0)$ and $\chi '(0)$, obtained 
by subtracting the ferromagnetic contributions from $S(0)$ and $\chi(0)$.
The parameters obtained in this case are $D'_s=0.106 \pm 0.001$, $D'_\chi=0.34 
\pm 0.01$, $T'_s=5.1 \pm 0.2$, and $T'_\chi= 3.9 \pm 0.3$. The high-energy 
cutoffs $T'$ obtained in this way are now much closer to each other, and also
close to the value $T_0=4.5$ obtained in Ref.~\onlinecite{SSS} by fitting a 
scaling form to the real-space correlation function at $\beta=32$. Within 
the accuracy of the numerical procedures (which apart from statistical errors
also include effects from non-asymptotic contributions) they can be 
considered equal. We note that these parameters still should be viewed as 
`effective' or temperature-dependent, and their asymptotic zero temperature 
values may only be reached at $T/J < 0.01$ as is the case for the uniform 
susceptibility \cite{bethe1} and the correlation length \cite{bethe2}.

The temperature dependence of $\chi (q=0)$ is important in the context of 
systems of weakly coupled spin chains, where it determines the critical 
temperature for ordering into a three-dimensional Neel state.\cite{Schulz2} 
A recent calculation for KCuF$_{\rm 3}$ by Schulz,\cite{Schulz2} using our
numerical data for $\chi(q=0)$, is in good agreement with experimental 
results.

\section{Conclusions}

In this paper, we have used several numerical methods to study the 
dynamics of the spin-$1/2$ Heisenberg chain at intermediate temperatures 
(above and below $J$). In order to obtain the dynamic structure factor, 
we have combined high temperature expansions for the frequency moments 
with the recursion method, and quantum Monte Carlo simulations with the 
maximum entropy method. In some cases, the results of these methods have 
been compared with exact results for a system with 16 spins, as an 
additional check.

We find that at high temperatures the HTE + recursion method works very 
well even for dynamic quantities. Using the first six continued fraction 
coefficients for the relaxation function, we have obtained dynamic 
susceptibilities at various temperatures down to $T/J=0.5$. The calculated 
frequency-dependence of the structure factor at $k=\pi/2$ agrees remarkably 
well with the exact results on finite systems. This method is also sensitive 
to infrared singularities. Already at the level of the continued fractions 
themselves, that is before any numerical extrapolation is done, the presence 
or absence of singularities can be detected. We found that the behavior of 
the continued fractions for the local ($k$-integrated) susceptibilities 
changes qualitatively depending on whether the hyperfine couplings are 
vanishing or non-vanishing at $k=0$. However, when infrared singularities 
are present, their exact forms may not be fully captured by these methods. 
Since the diffusion-related singularities are most robust at infinite $T$, 
where they have been investigated in the past with higher number of moments, 
\cite{Bohm} we did not focus on this issue here. In our study, the HTE with 
the recursion method became unreliable below $T/J=0.5$ because the 
extrapolations for the higher moments became unstable. By extending the 
series for the frequency moments, it may be possible to reach still lower 
temperatures also for dynamic quantities.

The QMC results for the static quantities agree perfectly with the
high temperature expansions down to $T/J\approx 0.1-0.2$, below which
the HTE results become unreliable. Results obtained for the dynamic 
quantities with the QMC + Max-Ent method are in satisfactory (generally 
within 10-20 percent) agreement with HTE results above $T/J=0.5$ and, we 
believe, should have similar reliability down to much lower temperatures. 
For calculating local quantities such as $1/T_1$, and incorporating as much 
as possible the effects of infrared singularities from certain $k$-regions, 
it appears to be better to carry out the analytic continuation for all
individual momenta separately before performing the momentum sum, rather 
than carrying out the analytic continuation for momentum-integrated 
quantities. 

We have also presented several new results for the Heisenberg chain. These 
include quantitative estimates for the logarithmic temperature dependence 
of the static staggered susceptibility and structure factor, improved 
estimates for the temperature dependence of the spin-lattice relaxation 
rate with various choices of hyperfine couplings, as well as the full dynamic 
structure factor at intermediate temperatures. Our results clearly show the 
shift in the low frequency spectral weight from the diffusive modes near 
$k=0$ at high temperatures to the antiferromagnetic modes near $k=\pi$ at 
low temperatures. At low temperatures we observed the development of spectral
features that can be associated with the two-spinon continuum. Overall, our 
results at the lowest temperatures are in good agreement with with general 
expectation of these quantities from various analytical studies. Our results 
also allow us to reconcile the measurements of the spin-lattice relaxation 
rates in Sr$_2$CuO$_3$ with theoretical and numerical calculations for 
the Heisenberg model.

After completing this work, we became aware of a recent paper by
Fabricius, L\"ow and Stolze,\cite{fls} discussing exact diagonalization 
results for $S(k,\omega)$ of chains with up to 16 sites. We have here used 
such short-chain calculations mainly as a test of the HTE and QMC + Max-Ent
methods in some regimes. Fabricius {\it et al.}\cite{fls} found no signs
of spin diffusion. Our results indicate that 16 sites is not large 
enough in general, in particular not for addressing the intricate problem 
of spin diffusion. As can be inferred from Fig.~\ref{s0qmc}, the momentum 
cut-off $k_1 = \pi/8$ for a 16-site chain prohibits access to most of the 
long-wavelength regime where we see signs of diffusive behavior. In the 
limit $T \to 0$ there is most likely no spin diffusion, as discussed
by Sachdev,\cite{Sachdev} and also supported by the numerical results of
Ref.~\onlinecite{Sandvik2} and our present calculations. We stress again 
that our study is also not accurate enough to resolve the exact form of 
$S(k,\omega)$ in the $k \to 0, \omega \to 0$ limit. We have strong evidence 
of a sharp peak at high temperatures, but cannot determine whether or not 
it is truly divergent (i.e., whether the long-time behavior of the spin-spin 
correlations is of the standard 1D diffusive form $\sim t^{-1/2}$, or, 
perhaps, is anomalous). In view of the absence of diffusion in the limit 
$T \to 0$, a sharp $\omega \to 0$ peak in $S_0(\omega)$ at high $T$, 
diverging as $T\to \infty$, is perhaps the most likely scenario. In any 
case, as we have discussed above, a sharp maximum should have detectable
effects on, e.g., the spin lattice relaxation rate in real materials. 

We would like to thank M. Horvatic, S. Sachdev, D. Scalapino, H. Schulz, 
and M. Takigawa for stimulating communication. This work was supported by NSF 
under Grants DMR-9318537 and DMR-9520776, and by the Campus Laboratory 
Collaboration of the University of California. The QMC calculations were 
carried out at the Supercomputations Research Institute at Florida State 
University.

\begin{figure}
\caption{Continued fraction coefficients, $\Delta_k$ for $S_R(\omega)$
as a function of $\beta$ for $R=0$ (lower panel) and as a function of the
index $k$ for $R=0$ and $-0.5$ at $\beta=0$ (upper panel).}
\label{coeff}
\end{figure}

\begin{figure}
\caption{Temperature dependence of the parameters of the fit
for two values of $R$.}
\label{fit}
\end{figure}

\begin{figure}
\caption{Comparison of the real and model $\Delta$-sequences
for two values of $R$ at different temperatures. Each panel
shows $\Delta_3$ (bottom), $\Delta_4$ (middle), and $\Delta_5$ (top).}
\label{delta}
\end{figure}

\begin{figure}
\caption{Comparison of exact $N=16$ (histograms), HTE (solid curves) and 
N=128 QMC + Max-Ent (dashed curves) results for the dynamic structure 
factor at $k=\pi/2$.}
\label{comp}
\end{figure}

\begin{figure}
\caption{The dynamic structure factor $S_R(\omega)$ with $R=0$
calculated using the HTE method at various temperatures.}
\label{swhte}
\end{figure}

\begin{figure}
\caption{The dynamic structure factor averaged with a constant form factor
calculated using QMC and Max-Ent analytic continuation at different
temperatures. Dashed and solid lines are results for system sizes
$N=64$ and $128$, respectively.}
\label{swqmc}
\end{figure}

\begin{figure}
\caption{The low-frequency limit of the dynamic structure factor vs.~the
momentum. Solid and open circles are for $N=64$ and $128$, respectively.}
\label{s0qmc}
\end{figure}

\begin{figure}
\caption{QMC + Max-Ent results for the full dynamic structure factor 
$S(k,\omega)$ at three different temperatures for $N=128$. The maxima of 
the vertical scales are $3.01$ ($T/J=1.0$), $1.49$ ($T/J=0.5$), and $1.86$ 
($T/J=0.25$).}
\label{sqw3d}
\end{figure}

\begin{figure}
\caption{QMC + Max-Ent results for the dynamic structure factor 
$S(k,\omega)$ at $T/J=1.0$, $0.5$, and $0.25$, with shades of grey
representing the intensity in the $(\omega,q)$ plane. The curves indicate
the lower and upper bounds at $T=0$. Note that for $T=1.0$ and $0.5$,
$S(k,\omega)$ is sharply peaked at $k \to 0, \omega \to 0$.}
\label{sqw2d}
\end{figure}

\begin{figure}
\caption{QMC + Max-Ent results for the dynamic structure factor at $k=3\pi/4$,
calculated for a $64$-site system at inverse temperature $\beta=24$. The
dashed line is the $T=0$ Bethe ansatz lower edge.}
\label{s34pi}
\end{figure}

\begin{figure}
\caption{Comparisons of $S_0(\omega)$ calculated by averaging before
(dashed curves) and after (solid curves) analytic continuation.}
\label{comps0}
\end{figure}

\begin{figure}
\caption{Results for NMR relaxation rates $1/T_1$ for various values of 
the hyperfine parameter $R$, calculated using HTE and the QMC and Max-Ent 
methods.}
\label{t1fig}
\end{figure}

\begin{figure}
\caption{The equal-time structure factor $S(q)$ obtained from HTE and QMC,
graphed so that the data would collapse onto a single curve if scaling 
in $q/T$ holds. The QMC results for $\beta \le 8$ were calculated in the
grand canonical ensemble for $N=256$, and those for $\beta \ge 16$ in
the canonical ensemble for $N=1024$.}
\label{sqfig}
\end{figure}

\begin{figure}
\caption{The $T$-dependence of $S(k=\pi)$ and $\chi(k=\pi)$ obtained 
from HTE and QMC, graphed so that the predicted forms 
(\protect{\ref{s0x0}}) give linear behavior. The dashed curves are several 
different HTE approximant, and the solid circles with error bars are the 
QMC results. The solid lines are fits to the QMC data for $T/J < 0.25$.
The open circles and dotted lines are results and fits after subtraction of 
the ferromagnetic contribution to the spin-spin correlation
function.}
\label{sxpi}
\end{figure}


\begin{references}
\bibitem{cnote}
In some cases it may be necessary to use the full covariance matrix
in Eq.~(\protect{\ref{chi2}}), instead of just its diagonal elements.
\protect{\cite{maxent1,maxent2}} However, for the problem under
consideration here, we find no significant differences between the
two approaches.
\end{references}
\end{document}